\documentclass{basi}
\usepackage{graphicx}

\begin{document}
\title[Expansion of double radio sources]
{Expansion of radio galaxies in a cosmologically
evolving medium: Possible implications for the
cosmic star-formation history}

\author[Barai et al.]%
       {Paramita Barai,$^1$\thanks{e-mail:barai@chara.gsu.edu}
        Gopal-Krishna,$^2$ M.\ Angela Osterman$^1$ and Paul J.\ Wiita$^1$ \\
	$^1$Department of Physics \& Astronomy, P.O. Box 4106, Georgia State University,
            Atlanta, GA\\ \hskip 1.0 true cm 30302-4106, USA \\
	$^2$National Centre for Radio Astrophysics/TIFR, Post Bag No.\ 3, 
            Pune University Campus,\\ \hskip 1.0 true cm Pune 411 007, India}





\maketitle
\label{firstpage}
\begin{abstract}
We compare earlier estimates of the volumes filled by lobes of radio
galaxies during the quasar era based upon non-evolving ambient media
with new ones assuming a strong cosmological evolution of the ambient
medium. 
If the sources remain active for
over $10^8$ years the volumes filled by them are found to be comparable 
for the two
scenarios. This strengthens our earlier inference that much of the 
cosmic web of gaseous filaments, the site of galaxy formation, was probably 
permeated by radio lobes during the quasar era and this could have 
triggered extensive star formation and made large contributions to the 
spread of magnetic fields and metals through the universe by $z \sim 2$.
\end{abstract}

\begin{keywords}
galaxies: active -- galaxies: jets  --
quasars: general -- radio continuum: galaxies -- stars: formation
\end{keywords}


\section{Introduction}
\label{sec:intro}
 A steady decrease with redshift, $z$, of the physical size,
 $D$, of radio luminous quasars was first noted over three 
 decades ago; with the parameterization $D \propto (1 + z)^{-n}$, 
 $n \sim 1$ was found
 (Legg, 1970; Miley, 1971).  This same result was generalized by
 Kapahi (1975) to double-lobed radio sources of Fanaroff-Riley type II 
 (Fanaroff \& Riley, 1974), using the angular size--flux
 density plot derived from the Ooty lunar occultation survey 
 (Swarup, 1975).  Subsequent studies using deeper radio
 surveys  indicated a 
 steeper cosmic evolution, with $n \sim 2$ out to $z \sim 0.5$
(Kapahi, 1985) or even as high as $n \sim 3.5$
(Oort et al., 1987). 
 The cosmological evolution of linear size was first interpreted in 
terms of a systematic decrease in the ambient  
density due to the expansion of the universe ($\rho \propto
 (1 + z)^{3})$ (e.g., Wardle \& Miley, 1974). Following the
 discovery of X-ray emitting hot gaseous halos around 
 massive elliptical galaxies, Baldwin (1982) considered a 
 more realistic, power-law density profile of the ambient 
 medium for the propagation of the jets. 

The evolution of 
 conical jets passing from such a power-law density distribution 
 of the interstellar medium (ISM), and then into a
 cosmologically evolving intra-cluster medium (ICM), after 
 crossing an ISM/ICM interface, was first considered analytically 
by Gopal-Krishna \& Wiita (1987) and numerically by Rosen \& Wiita 
(1988) and  Wiita et al.\ (1990). These models were 
indeed able to account for the observed steep size--redshift 
evolution. However, were this outer confining gas to be identified with an
all pervasive intergalactic medium (IGM), its assumed properties
and redshift
dependence would be inconsistent with limits on the
Compton $y$ parameter (e.g., Rosen \& Wiita, 1991) set by the COBE 
microwave background radiation measurements (Smoot et al., 1991). 

A subsequent  detailed analytical study
also took into account the increase in the ``radiative
 efficiency'', 
$\epsilon$, of the synchrotron radio lobes surrounded
 by a denser medium (Gopal-Krishna \& Wiita, 1991).
There we showed 
 that nearly half of the steep linear size evolution could be 
 attributed to this 
$\epsilon$ factor, and the remainder 
 could be explained in terms of a cosmological evolution of 
 the King-type density profile of the hot gaseous halos of 
 the massive ellipticals.   However, later studies of
the redshift--size--power relations indicated a somewhat weaker
dependence of size on redshift (e.g., Neeser et al., 1995).
In a radical departure from
 this approach, Blundell \& Rawlings (1999) more recently argued
 that the observed linear size evolution may well be an artefact of the 
``youth--redshift degeneracy'' of radio sources, 
which is imposed by the steeply rising inverse Compton losses 
against the cosmic microwave background at earlier epochs
(see Rees \& Setti, 1968; Gopal-Krishna et al., 1989),
coupled with substantial adiabatic losses as the lobes expand.  
 
Several of  these interpretations rest on the simplifying assumptions 
that the shape and central density of the radial density profiles 
of the gaseous halos do not evolve with the cosmic epoch. 
In fact, Blundell et al.\ (1999, hereafter BRW)
have claimed that there is no good evidence for a changing environment 
of most radio galaxies (e.g., Mulchaey \& Zabludoff, 1998), 
even back to quite early epochs $(z \sim 3)$.
Nonetheless, several lines of observational evidence
indicate changes in radio galaxy environment with redshift. 
At medium redshifts ($z \sim 0.5$), moderate to high power radio galaxies 
are found in environments richer by $\sim 2 - 3$ times, 
compared to their counterparts in the nearby universe ($z < 0.1$)
(Hill \& Lilly, 1991; also, Yates, Miller \& Peacock, 1989).
More distant $(z \sim 1)$ radio galaxies in the 3CR catalog are more
frequently associated with moderately rich (proto-) clusters, whereas the nearby 
3CR FR II galaxies are usually found in more sparsely populated regions, 
implying that the cluster environments of the 3CR galaxies 
have changed with redshift (Best et al., 1998).

Good fits 
to the radio power, $P$, linear size, $D$, 
and $z$ distributions of complete samples of radio galaxies selected
at meter wavelengths were claimed by BRW, who assumed a 
non-evolving ambient medium and a constant radius for the hot-spots.
Empirically derived radio source 
properties include a large active lifetime of the central engine, $\sim
5 \times 10^8$yr (e.g., BRW; Barger et al., 2001; McLure \& Dunlop, 2004) and 
a power-law
beam-power function with a slope of about -2.6, based on matching complete samples of radio sources
with essentially complete redshift data (BRW).  

Employing the above results and assumptions, Gopal-Krishna \& Wiita (2001, 
hereafter GKW01) showed that during the `quasar era' ($z \sim 2-3)$, 
much of the denser (proto-galactic) material in the universe (which was 
concentrated within the cosmic sheets and filaments) was directly impacted
by the expanding lobes of the generations of radio galaxies born during
that era.  Denser clumps of gas scattered across those
cosmic filaments could thus be compressed, yielding global starbursts,
and the overpressured radio lobes could trigger, or
at least, accelerate, the formation of entire new galaxies (GKW01; Gopal-Krishna
and Wiita, 2003a, hereafter GKW03a;
cf.\ Daly, 1990).
They further argued that this picture of a radio lobe-filled
 early universe could not only explain the much higher star 
formation rate found at high redshifts (e.g., Archibald et al., 2001) 
but also readily account for the presence of magnetic fields 
in distant galaxies (GKW01; Gopal-Krishna et al., 2003
hereafter GKWO). 
Two entirely independent groups, approaching the problem from
different directions, also concluded that QSOs were 
energetically capable of 
magnetizing the universe
(Kronberg et al., 2001; Furlanetto \& Loeb, 2001).
Our scenario can also account for the widespread distribution of 
metals in the proto-galaxies seen at these high redshifts 
(Gopal-Krishna \& Wiita, 2003b, hereafter GKW03b).

In this communication, we wish to revisit this issue by 
relaxing the simplifying assumption of a redshift-independent 
environment of powerful radio galaxies. Accordingly, we present 
a simple quantitative treatment of the beam propagation, 
roughly taking
into account cosmic evolution of both the density and radial
distribution of gas around  radio-loud ellipticals. It is expected
that this first-order analytic treatment will inspire detailed 
numerical simulations of this important problem.

\section{Beam propagation in a cosmologically evolving medium}

 Recent studies of structure formation in the universe
 have led to a picture wherein the formation of stars and 
 galaxies occurs inside a cosmic web of filaments containing
 the proto-galactic baryonic material 
(e.g., Cen et al., 2001). From these $\Lambda$CDM 
simulations, it is estimated that during the quasar era ($z 
\sim 2 - 3)$, the 
filament 
contained $\sim 10\%$ of the total baryons but 
occupied only $\sim 3\%$ of the volume of the universe 
(e.g., Cen \& Ostriker, 1999; Dav{\'e} et al., 2001).
 The corresponding numbers for the present epoch are estimated 
 to be $\sim 50\%$ and $\sim 10\%$, respectively. Taken at face 
 value, these estimates, together with the growth factor of the 
 volume of the universe, $(1 + z)^3$, suggest that the {\it mean} 
 gas density in the filaments during the quasar era was roughly 
 a factor of 30 (i.e., $\left[(0.1/0.5)(0.1/0.03)\right] [1+2.5]^3$)  
higher than at the present epoch.  This very different environment
for RGs during that epoch is also indicated by the evidence
cited above for intermediate $z$ (Yates et al., 1991; Hill \& Lilly, 1991;
Best et al., 1998).  
We next consider the radial distribution of the ambient 
 gas around the elliptical galaxy hosts of double radio sources. 
 In all our computations of the volume attained by the radio 
 lobes (GKW01), we have adopted a power-law (essentially King model)
 density profile for the
 hot gaseous halos: 
$n(r) = n_0(r/a_0)^{-\beta}$, 
with  $n_0 = 1.0 
\times 10^{-2}{\rm cm}^{-3}$, $a_0 = 10~$kpc, and
$\beta = 1.5$, which is also used by BRW and is based on  X-ray
 images of nearby massive ellipticals in groups (e.g., Sarazin, 1988;
Mulchaey \& Zabludoff, 1998). 
 While such a typical radial density distribution is certainly appropriate for
 small redshifts, this may not to be a good approximation at 
 the high redshifts corresponding to the quasar era, which 
 witnessed a $10^2 - 10^3$ times higher co-moving density of powerful radio-loud 
 ellipticals (e.g., Jackson \& Wall, 1999). At such early 
 epochs, the cosmic filaments were accreting gas vigorously. 
The hot gas is likely to have been  more uniformly 
 distributed within the filaments, but was in the process of becoming 
 increasingly non-uniform due to gravitational accretion onto
 the dark matter halos and galaxies existing or forming within 
the filaments (e.g., Cen et al., 2001;
Viel et al., 2003).  

For the gas distribution around the elliptical hosts at $z = 2 - 3$, 
we consider two models. Model 1 adopts the mean density distribution
around the massive ellipticals found in the local universe (which has
 an essentially power-law profile; see above).  However, 
given the lack of any reliable estimate for the degree of non-uniformity 
of the medium which had developed by such early epochs, we shall 
consider an alternative simplifying assumption as well. 
Thus, in our model 2, at $z \sim 2 - 3$, the 
galaxies within 
 the filaments are taken to be immersed in an ambient gas of uniform density. 
These two gas density profiles adopted by us for
the quasar era represent two extreme situations.
 The uniform density (for model 2) is $N (\sim 30)$ times higher 
 than the mean density of the ambient gas in model 1, 
which was used in 
 our previous analytical work (GKW01; GKW03a,b; GKWO), 
 in which we had adopted an ambient
 density profile for $z = 2.5$ ellipticals identical to that 
 found for the nearby massive ellipticals (and consistent with 
 the inference of BRW).

To derive the mean gas density, we consider the mass of ambient gas 
contained within a sphere of radius $R$. 
From the mean density ratio ($N$) mentioned above, 
the mass within the spherical volume 
in the constant density case (model 2)
is $N$ times larger 
than in model 1, the case with a power-law density decline with radius. 
We average over the spherical volumes 
to calculate the mass contained in the sphere for model 1,
and then after multiplying with the factor $N$, obtain the values of 
the constant density, $\rho_{2}$, to be used in model 2.
A few representative radii, $R = 3, 5$, and $10$ Mpc are considered.

The temporal 
 growth of the total extent of a radio source, assuming a cylindrical
beam, as is roughly appropriate for large, powerful sources (e.g., Jeyakumar
\& Saikia, 2000) can be computed, and the total length of 
the cigar shaped radio cocoon for 
model 1 is given by
(e.g., Kaiser et al., 1997)
\begin{equation}
D_1(t) = 3.6 a_0 \left(\frac{t^3 Q_0}{a_0^5 \rho_0}\right)^{1/(5-\beta)}.
\end{equation}
The cocoon length for the constant density case, model 2, is
 \begin{equation}
     D_2(t) = 3.6 \left(\frac{t^3 Q_0}{\rho_2}\right)^{1/5}.
\end{equation}
Here $\rho_0 = m_p n_0$, $t$ is the time for which the source has been expanding, 
$Q_0$ is the 
(assumed constant) power fed into each jet, and $m_p$ is the mass of a proton.
From the previous discussion 
we find that
$\rho_2 = [3N \rho_0/(3-\beta)] (a_0/R)^{\beta}$.
We define

\begin{equation}
     \eta(t) \equiv \frac{D_1(t)}{D_2(t)} =  \left(\frac{t^3 Q_0}{a_0^5}\right)^{\beta/[5(5-\beta)]}
   \left(\frac{\rho_2^{1/5}}{\rho_0^{1/(5-\beta)}}\right).     
\end{equation}
 
For the two extreme scenarios, Fig.\ 1 shows 
 the ratio, $\eta$, of the sizes attained at different ages
 of the radio source,
taking $R = 5$ Mpc and considering several values of $N$.  
Even allowing for a wide range in
both $N$ and $R$ we find that $0.2 < \eta < 2.0$. 
The total volumes filled by the lobes, for fixed $N$ and $R$, but for
three beam powers spanning most of the relevant range,
are displayed in Fig.\ 2.  These volumes were computed using
the method of BRW (their Eq.\ 20). We see that for either model,
volumes approaching or exceeding 1 Mpc$^3$ are found for all
FR II radio source powers if the sources are (quasi-) continuously
active for times exceeding $10^8$ yr; the curves for $Q_0 = 5 \times 10^{44}$
erg s$^{-1}$ correspond to the weakest such sources (BRW; GKW01).
Such large lobe volumes, when integrated over appropriate
distributions in $Q_0$ and $z$ (GKW01, GKW03a), imply that the radio lobes would
fill substantial fractions of the cosmic web during the quasar era. 

\begin{figure}

\resizebox{\hsize}{!}{\includegraphics{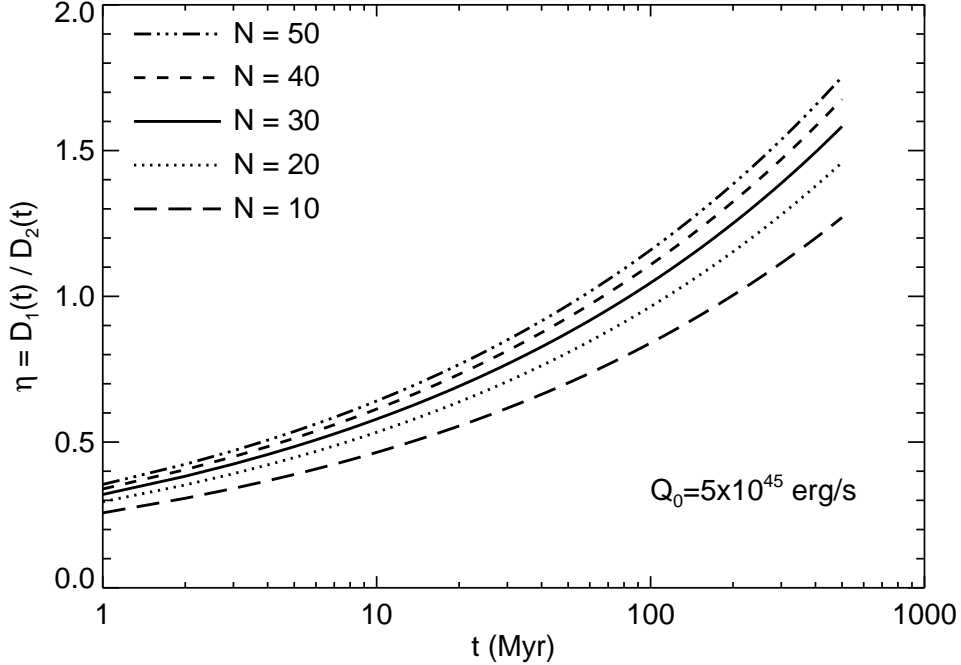}}
\caption{Distance ratios for the power-law model 1 to
the enhanced constant density model 2 against source age for
a typical FR II power, $Q_0$ and for 
several density enhancement factors, $N$; $R = 5$ Mpc is assumed.}
\label{fig1}
\end{figure}

\begin{figure}
\resizebox{\hsize}{!}{\includegraphics{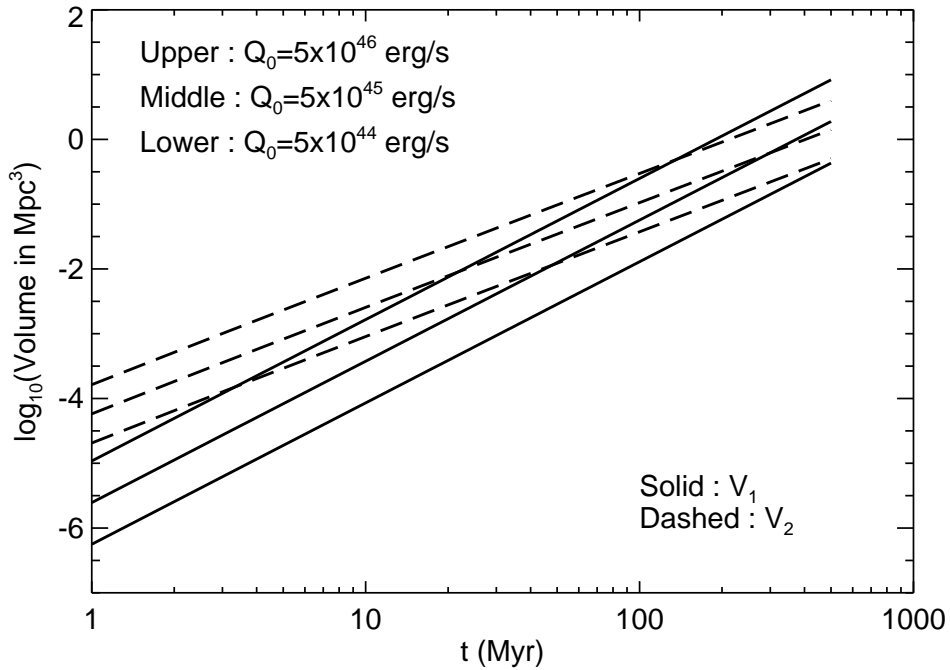}}
\caption{Total volumes filled by the lobes for the two different
models as functions of time for several beam powers; $N = 30$ 
and $R = 5$ Mpc are assumed for model 2.}
\label{fig2}
\end{figure}

\begin{figure}
\resizebox{\hsize}{!}{\includegraphics{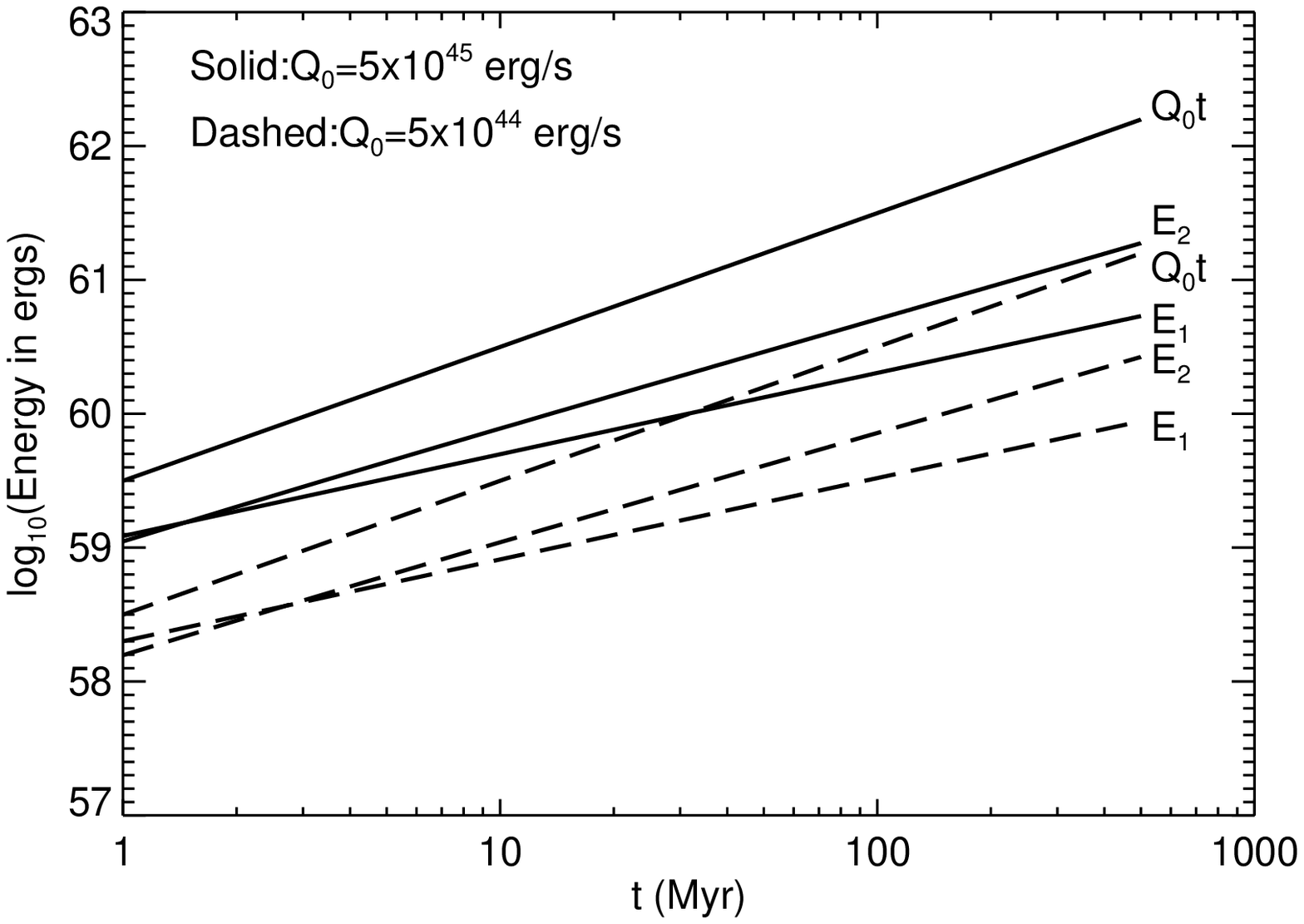}}
\caption{Injected energies ($Q_0 t$) and the energies stored in the
lobes ($E$) at different ages, for the two models. $N = 30$ and $R = 5$ Mpc are adopted
for model 2.}
\label{fig3}
\end{figure}

The total energy  inputs into the radio lobes and the surrounding
medium, $Q_0 t$,  as well as the pressure-weighted volumes of the lobes,
corresponding to the energies stored in the lobes, $E$ 
(computed using  Eqs.\ (9) and (20) of BRW), for the
 two models of ambient density profile are displayed in Fig.\ 3. 
This energy stored in the lobes  has  
 a more direct bearing on the efficacy of starbursts being
 triggered by the collapse of the 
denser and cooler 
gaseous clumps engulfed
 by the lobe.  The comparable values of $E_1$ and $E_2$
(never different by more than a factor of 3) for the two models imply that the 
impact of
the bow shocks on any clouds within the IGM will be similar
under both of these scenarios.  These results are not dramatically affected
when we considered different values of $N$ between 10 and 50
and $R$ between 3 and 10 Mpc; those parameter values should roughly span 
the conditions prevailing during the quasar era.

\section{Discussion and Conclusions}

A rather unexpected result emerging from the above computation is
that even if the average ambient density through which
the lobes must penetrate is much
greater during the quasar era, RGs can still grow to enormous
sizes. The volumes that they can fill, while still
significantly overpressured, can easily exceed 1 Mpc$^3$.  
Furthermore, the total energy available to compress gaseous clouds
in the IGM, and thereby trigger extensive star- and even galaxy-formation,
is comparable for the models where jets propagate through
unevolving galactic
halos  and those which assume propagation through 
a uniform ambient medium with mean density scaling roughly
as $(1+z)^3$.  Recent hydrodynamical simulations which include
cooling of lobe-type
shocks interacting with large clouds clearly reveal
the formation of 
numerous dense cooling clumps which should 
continue to collapse into star clusters (Mellema et al., 2002;
Fragile et al., 2004a,b). 
Observational support for this hypothesis comes from
the evidence for a young stellar component 
in the extended optical emission 
revealed by the HST observations of high$-z$ RGs 
(e.g. Best et al., 1996; Dey et al., 1997; Bicknell et al., 2000).
In view of the above, our earlier inference concerning the
role of radio galaxies in the cosmic star formation history 
(GKW01, GKW03b, GKWO) appears to hold considerable promise
and needs further invstigation.

To summarize, our earlier key conclusion, that RG lobes could 
accelerate or even trigger the formation of many new galaxies 
during the quasar era (GKW01), appears to be fairly robust.  In 
addition, the infusion of
magnetic fields of significant strengths ($\sim 10^{-8}$ G, e.g.,
Ryu et al., 1998)
into at least the cosmic web portion of the IGM certainly
could have been caused by the lobes of radio galaxies
(GKW01; Furlanetto \& Loeb, 2001; Kronberg et al., 2001;
GKWO).
An alternative picture of magnetising the IGM comes from the 
superwinds driven by outflows from stars and galaxies 
(Kronberg et al., 1999). However, this situation would not 
naturally lead to 
preferential alignment between radio lobes and newly forming galaxies 
as has been observed (e.g., Best et al., 1996; Bicknell et al., 2000). 
Also, the radio loud AGNs outside clusters offer a potentially 
more energetic route for magnetisation of the wider IGM 
(Kronberg et al., 2001)

Finally, these expanding radio lobes could
have swept out the metal-rich ISM of young galaxies they encounter
(including that of the active hosts), thereby contributing substantially
 to the widespread metal pollution of the IGM (GKW03b).
Recent evidence of super-solar metalicities in quasar
nuclei at $z \sim 4$ (e.g., Dietrich et al., 2003) 
and substantial metalicity in
even underdense regions of the IGM at similarly high $z$
(e.g., Schaye et al., 2003) strongly hints at the need for an 
efficient means of spreading metals widely at early cosmic
epochs.  While the obvious sources for the production of metals detected in
quasars are starbursts in the host galaxies, the possibility of
nucleosynthesis in the accretion disks feeding the central
black holes  (e.g., Mukhopadhyay \& Chakrabarti, 2000; 
Kundt, 2002) also should be considered.
It is possible 
that the superwind outflow model (Kronberg et al., 1999) would also 
contribute to the metalization of IGM, though that aspect
of this scenario has not yet been investigated.

Here, we draw attention to a recent work by Rawlings \& Jarvis
(2004). While agreeing that RG lobes will penetrate much of the 
universe, they argue that this may often shut off star formation
by expelling gas from protoclusters. However, unlike our picture
(see, also, Rees 1989), they assume a single phase medium. 

To test our picture, observation of Giant Radio Galaxies 
at low frequencies using the Giant Metrewave Radio Telescope (GMRT) 
(e.g., Konar et al., 2003), can be useful. Still more vital input
is expected from upcoming low frequency radio telescope LOFAR (Low 
Frequency ARray), which holds the potential of observing high$-z$ radio 
galaxies and, especially, the fading
giant RG's at $z > 1$ (R{\"o}ttgering, 2003).

In future work we will explore more sophisticated models where the
jets first propagate through halos with evolving central densities
and eventually enter the IGM, whose density will also be
allowed to evolve.  Models where the jets are  conical 
(cf.\ Gopal-Krishna \& Wiita, 1987, 1991), at least
at early times (e.g., Jeyakumar \& Saikia, 2000) will also 
be considered.  In addition, we will compare the
resulting distributions of RG sizes, powers and redshifts against
observational surveys so as to simultaneously constrain the 
range of evolutionary models for both the IGM and the radio source
population.

\section*{Acknowledgements}
We thank P.\ Kronberg, J.\ P.\ Leahy
and J.\ P.\ Ostriker for useful conversations and the
referee for helpful suggestions. PJW is grateful for
hospitality at the Princeton University 
Department of Astrophysical Sciences.
PB, MAO and PJW are partially supported by Research Program Enhancement
 funds to the Program in ExtraGalactic Astronomy
at GSU.


\label{lastpage}
\end{document}